\magnification=\magstep1
\advance\voffset by 3truemm
\advance\hoffset by -1truemm
\vsize=23truecm
\hsize=16.5truecm
\overfullrule=0pt
\hfuzz 5truept
\parskip=5pt
\baselineskip=12pt
\font\abstractfont=cmr9
\font\abstracttitlefont=cmti9
\font\authorfont=cmr10 scaled\magstep1
\font\sectionfont=cmbx10 scaled\magstep2
\font\titlefont=cmbx10 scaled\magstep2
\def\title#1{\null\vskip26truemm\noindent{\titlefont#1}}
\def\author#1{\vskip15truemm\noindent{\authorfont#1}}
\def\address#1{\vskip5truemm\noindent#1}
\def\abstract#1{\vskip26truemm\noindent{\abstracttitlefont Abstract.}
{\abstractfont#1}\vskip16truemm}
\newcount\numsection
\def\references{\bigskip\bigbreak\noindent\leftline{\bf REFERENCES}\bigskip}
\def\section#1{\advance\numsection by1
\bigskip\bigbreak\noindent\leftline{\sectionfont\the\numsection\
#1}\bigskip\nobreak}
%
%

\let\epsilon=\varepsilon

\def\frac#1#2{\hbox{$#1\over#2$}}

\title{INTRODUCTION \`A LA PHYSIQUE QUANTIQUE}
\author{C. Piron}
\address{D\'epartement de Physique Th\'eorique, 24 quai Ernest-Ansermet,
CH-1211 Gen\`eve 4}
\abstract{En nous laissant guider par la notion de champ qui en fait
domine toute la physique nous d\'efinissons la nature physique d'un
syst\`eme, ses propri\'et\'es et ses \'etats possibles. Nous proposons
alors un cadre tr\`es g\'en\'erale permettant la description de tels
syst\`emes et la  construction de mod\`eles consistants.
Nous en donnons des exemples et pour illustrer les concepts de notre
th\'eorie nous d\'ecrivons  diff\'erents
types d'\'evolutions. Pour terminer nous discutons en d\'etails un
processus physique, les \'echos de spins.}
\section{La nature d'un syst\`eme physique}
Pour comprendre l'\'evolution de notre conception de la physique en y
int\'egrant ses aspects quantiques sans en \^etre choqu\'e ou
boulevers\'e, il faut absolument abandonner la vieille image d'un monde
form\'e de tr\`es petites particules de mati\`ere, circulant dans un vide
qui n'aurait aucune existence en soi. Au contraire, il faut admettre comme
fondamental l'existence de l'espace vide et du temps qui
s'\'ecoule [Barut 1993]. Les particules  ne sont
alors rien d'autre que des manifestations de l'espace,
des sortes de singularit\'es du champ [Piron et Moore 1995]. Il peut
 para\^{\i}tre surprenant que l'espace et le temps
puissent \^etre trait\'es s\'epar\'ement, mais c'est le r\'esultat
qui s'impose quand on constate que les points de l'espace-temps
de Minkowski, d\'efinis par les satellites qui gravitent au voisinage
de la terre, restent toujours confin\'es dans un sous-espace \`a trois
dimensions correspondant \`a une m\^eme valeur du temps coordonn\'ee
[Piron 1990 et Audoin et Guinot 1998].
Ainsi contrairement au pr\'ejug\'e du si\`ecle pass\'e
il n'y a pas un  seul et m\^eme espace de Minkowski dans lequel nous
serions tous plong\'es, mais chaque observateur, ayant adopt\'e un rep\`ere
de Galil\'ee, construit avec ses unit\'es son propre espace-temps de
Minkowski \`a partir du m\^eme espace et du m\^eme temps physique.
A premi\`ere vue, il semble qu'il y aurait une difficult\'e conceptuelle
\`a attribuer des propri\'et\'es \`a l'espace vide, comme par exemple
 affirmer qu'il est quasi-Euclidien et qu'il y r\`egne  un champ de gravitation.
En effet comment v\'erifier de telles affirmations sans devoir y introduire
des appareils et dans ce cas on n'a
plus le vide. Cet apparent paradoxe  a \'et\'e r\'esolu par Dirk Aerts
 gr\^ace \`a une formulation pr\'ecise de la notion d'\'el\'ements de r\'ealit\'e
jointe \`a une d\'efinition pr\'ecise des projets exp\'erimentaux
[Aerts 1982].
En effet selon Aerts, un projet
exp\'erimental est une exp\'erience, qu'on pourrait fort bien
\'eventuellement ex\'ecuter, et
dont le r\'esultat positif a \'et\'e  d\'efini une fois pour
toutes.
En plein accord avec la d\'efinition d'Einstein, Aerts pr\'etend alors,
que le syst\`eme poss\`ede un
\'el\'ement de r\'ealit\'e et que la propri\'et\'e est actuelle, si
 on peut affirmer
 par avance que dans l'\'eventualit\'e de l'ex\'ecution du projet
 correspondant
la r\'eponse positive est certaine. Ainsi, affirmer qu'ici l'espace
 est Euclidien, c'est affirmer qu'ici la somme des angles d'un triangle
qu'on construirait \`a l'aide de trois solides r\`egles rectilignes serait
\`a coup s\^ur \'egale \`a 180$^\circ\,$. De m\^eme
affimer qu'ici r\`egne un champ de gravitation, c'est pr\'etendre
qu'une particule qui serait l\^ach\'ee ici serait
certainement acc\'el\'er\'ee vers le bas. Il est important de
remarquer qu'\^etre Euclidien et poss\`eder un champ de gravitation sont
des propri\'et\'es propres \`a l'espace, donc en l'absence
d'appareils. Enfin, par d\'efinition, se donner l'\'etat actuel de l'espace
(ou plus modestement de l'espace d'une cavit\'e), c'est se donner
toutes ses propri\'et\'es actuelles; autrement dit, c'est se donner
l'ensemble des exp\'eriences
 qui si on  r\'ealisait l'une quelconque d'entre-elles, celle-ci
r\'eussirait \`a coup s\^ur.
\section{Description g\'en\'erale d'un syst\`eme}
Dans [Piron 96] nous avons r\'eussi \`a d\'eriver  les r\`egles habituelles
de la th\'eorie quantique \`a partir de cette notion de champ, en
utilisant pour la dynamique un principe du m\^eme
type que celui qu'on
utilise pour formuler les champs de jauges
du mod\`ele standard. Nous ne voulons pas d\'eriver
ces r\`egles [Cohen-Tannoudji 1973 II A,1 et III B,1,2,3] ici mais
juste les rappeler, en
pr\'eciser les limites de validit\'e, en montrer les g\'en\'eralisations
et finalement en donner
l'interpr\'etation physique.
Donnons-nous un syst\`eme, c'est-\`a-dire, une partie limit\'ee de
la r\'ealit\'e physique, celle que nous voulons d\'ecrire et
qui est suffisamment prot\'eg\'ee du reste de l'univers pour \^etre alors
bien d\'efinie au cours du temps.
Admettons donc qu'un tel syst\`eme existe et puisse \^etre valablement
d\'ecrit par ses propri\'et\'es et ses
\'etats possibles, en plein accord avec les d\'efinitions et concepts
du paragraphe pr\'ec\'edent.
D'une mani\`ere g\'en\'erale, sur l'ensemble $\cal L$ de toutes les
propri\'et\'es de ce syst\`eme,
 nous d\'efinirons une relation
d'ordre: nous dirons que la propri\'et\'e $a$ est plus forte que
la propri\'et\'e $b$, ce que nous noterons $~a<b~$, si chaque fois que $a$
est actuelle $b$ est actuelle. De m\^eme, sur l'ensemble $\Sigma$
de tous les \'etats correspondants possibles nous d\'efinirons une
relation d'orthogonalit\'e: nous dirons que deux
\'etats ${\cal E}_1$ et ${\cal E}_2$ sont orthogonaux s'il existe un
projet exp\'erimental dont le r\'esultat positif est certain dans
l'\'etat ${\cal E}_1$ et impossible dans l'\'etat ${\cal E}_2$ .
En physique classique on remarque tout de suite que deux \'etats sont
orthogonaux d\`es qu'ils sont diff\'erents.
 Il est commode de repr\'esenter une propri\'et\'e
 $a\in{\cal L}$ par le sous-ensemble $A$ des \'etats de $\Sigma$
pour lesquels la propri\'et\'e $a$ est actuelle. Le sous-ensemble
des \'etats orthogonaux \`a tous ceux d'un $A$ donn\'e est
not\'e $A^\perp$.
A l'aide de quelques axiomes tr\`es g\'en\'eraux [Piron 1998 1.7.2]
on peut alors
d\'emontrer  que $A^\perp$ est une propri\'et\'e, l'orthogonale de $A$,
et que les sous-ensembles de $\Sigma$ qui repr\'esentent des propri\'et\'es
sont exactement les sous-ensembles biorthogonaux [Piron 1998 1.7.3].
Une propri\'et\'e repr\'esent\'ee par le sous-ensemble $A$ est dite
classique si
son orthogonale est repr\'esent\'ee par le
compl\'ementaire de $A$. Dans la suite nous conviendrons de
 ne plus distinguer une propri\'et\'e $a$ de son sous-ensemble
repr\'esentatif $A$. Deux \'etats ${\cal E}_1$ et ${\cal E}_2$ sont dit
macroscopiquement \'equivalents si les propri\'et\'es classiques
ne permettent pas \`a elles-seules de les distinguer en d'autres
termes, si chaque
propri\'et\'e classique qui contient l'un des deux \'etats contient aussi
l'autre. Les
classes d'\'equivalences ainsi d\'efinies sont appel\'ees les \'etats
macroscopiques. Ces classes  forment un ensemble dont les
sous-ensembles cor\-res\-pondent univoquement aux propri\'et\'es classiques
du syst\`eme.
On voit ainsi que la description classique d'un syst\`eme doit
\^etre consid\'er\'ee comme une premi\`ere approche
o\`u on n'aurait consid\'er\'e que les propri\'et\'es classiques
et  n\'eglig\'e les autres. Ainsi le formalisme d\'evelopp\'e \`a Gen\`eve durant
ces quarante derni\`eres  ann\'ees permet tout naturellement d'englober,
\`a la fois dans une m\^eme th\'eorie, les aspects
classiques et quantiques d'un syst\`eme physique. On remarquera que,
dans cette th\'eorie, l'aspect classique ne provient pas d'un brouillage
d\^u \`a l'influence d'un ext\'erieur al\'eatoire mais qu'au contraire
c'est des conditions nettes et pr\'ecises  qui permettent au
syst\`eme d'acqu\'erir les propri\'et\'es quantiques que
des exp\'eriences fines vont alors mettre en \'evidence.
Donnons maintenant trois exemples:
1) Les \'etats d'une particule classique sont d\'ecrits par les points
 $(\vec x,\vec p,t)$ de $R^7$. Deux \'etats sont physiquement orthogonaux
d\`es qu'ils sont diff\'erents. Ainsi chaque
sous-ensemble de $R^7$ est biorthogonal et repr\'esente
une propri\'et\'e. Chaque point de $R^7$ est \`a lui seul un \'etat
macroscopique.
C'est le prototype d'une th\'eorie m\'ecanique. Longtemps
certains physiciens, plus ou moins sectaires, se sont acharn\'es \`a
 vouloir tout expliquer, m\^eme la physique quantique, dans ce sch\'ema
purement m\'ecanique.
2) Les \'etats d'une particule quantique sans spin sont d\'ecrits
par les rayons d\'efinis par les fonctions d'un m\^eme espace de Hilbert,
l'espace $L^2(R^3,dv)$. L' orthogonalit\'e physique que nous venons de
d\'efinir
  est alors identique \`a l'orthogonalit\'e g\'eom\'etrique et
les sous-ensembles biorthogonaux sont exactement les sous-espaces
lin\'eaires ferm\'es. Il n'y a qu'un \'etat macroscopique, l'espace
tout entier.
C'est le prototype d'une th\'eorie hilbertienne. D'autres
physiciens, tout aussi sectaires, s'acharnent alors \`a vouloir tout
expliquer, m\^eme  la physique classique, dans ce sch\'ema purement
hilbertien.
Bien que toute variable classique soit manifestement
exclue par cette
description, physiquement il y en a au moins une, c'est la
variable $t$ dont la valeur permet de pr\'eciser
l'instant actuel du syst\`eme. La particule quantique doit donc
 en fait \^etre d\'ecrite par une famille d'espaces de Hilbert
index\'es par $t$. Des \'etats \`a deux temps diff\'erents ne peuvent
pas \^etre superpos\'es, car ils sont macroscopiquement s\'epar\'es.
On peut montrer rigoureusement qu'il est parfaitement impossible
de plonger toute cette
famille  dans un seul espace de Hilbert, car ce dernier n'a qu'un nombre
d\'enombrable de sous-espaces orthogonaux entre-eux et non un
continuum. Mais ce n'est pas tout il y a encore d'autres variables
classiques, moins explicites, mais qui doivent aussi
entrer  dans cette description comme par exemples celles qui
 d\'efinissent la masse, la charge
 ou encore le rep\`ere de Galil\'ee choisi.
Ainsi, consid\`erons le centre de gravit\'e:
  $$X^j=\int\phi_t^\ast(\vec x)x^j\phi_t(\vec x)dv$$
et la quantit\'e de mouvement totale:
  $$P_j=\int\phi_t^\ast(\vec x)(-i\hbar\partial_j\,)\phi_t(\vec x)dv$$
Ce sont des grandeurs qui d\'ependent manifestement de ce choix. En
changeant de rep\`ere de Galil\'ee on change leurs valeurs.
 Gr\^ace au th\'eor\`eme d' Ehrenfest, on con\c coit alors
l'utilit\'e d'un mod\`ele o\`u la particule serait donn\'ee
simplement par une famille d'espaces de Hilbert index\'es par
 $(P,X,t)$ et o\`u chaque espace ne d\'ecrirait que la partie interne
correspondante de l'\'etat. C'est dans
ce cadre que nous avons formul\'e le mod\`ele \`a deux corps de l'atome
d'hydrog\`ene o\`u les \'etats stationnaires existent bel et bien
[Piron 1965, Piron 1998 5.7].
C'est aussi dans ce cadre qu'il faut comprendre le formalisme \`a deux
niveaux [Cohen-Tannoudji 1973 IV C] fr\'equemment utilis\'e dans les
calculs d'exp\'eriences.
 
3)Il est hors de doute  qu'un atome avant m\^eme d'interagir et
d'entrer dans une cavit\'e est
 s\'epar\'e de celle-ci. Ce sont deux syst\`emes s\'epar\'es car on a
encore la possibilit\'e \`a cet instant de faire
une exp\'erience sur l'un des deux sans perturber l'autre. Or si nous
consid\'erons deux tels syst\`emes s\'epar\'es,
d\'ecrits chacun par son espace de Hilbert, le syst\`eme global ne
peut pas \^etre d\'ecrit valablement et sans contradic\-tion
par le produit tensoriel des deux espaces,
c'est le paradoxe EPR [Einstein 1935]. Dans sa th\`ese, Dirk Aerts
a r\'esolu ce probl\`eme [Aerts 1982].
Les \'etats possibles du syst\`eme global
correspondent aux couples  $({\cal E}_1,{\cal E}_2)$, form\'es
d'un \'etat du premier et d'un \'etat du deuxi\`eme, c'est-\`a-dire
au produit cart\'esien. Dirk Aerts d\'emontre alors
que deux  \'etats du produit
 $({\cal E}_1,{\cal E}_2)$ et $({\cal F}_1,{\cal F}_2)$ sont physiquement
 orthogonaux si et seulement si ${\cal E}_1$ est orthogonal \`a
${\cal F}_1$ ou  ${\cal E}_2$ est orthogonal \`a ${\cal F}_2$. Il montre
ensuite que les
propri\'et\'es physiques du
syst\`eme global satisfont les axiomes g\'en\'eraux et donc
correspondent univoquement aux sous-ensembles
biorthogonaux de ce produit, mais ne correspondent pas seulement
aux sous-espaces lin\'eaires ferm\'es du produit tensoriel ou m\^eme
d'un autre espace de Hilbert [Piron 1998 1.8.2, 1.8.3 et 8.2.3]. Ainsi la
contradiction est lev\'ee et la suggestion d'Einstein faite \`a la fin
 de l'article de 1935 est parfaitement correcte.
Ce sont ces \'etats produits qui sont couramment
utilis\'es dans la litt\'erature (sans justification ni r\'ef\'erence!)
pour d\'ecrire  avant toute interaction les \'etats possibles d'un
syst\`eme compos\'e.
Tous ces exemples sont comme toujours rien plus que des mod\`eles, mais
c'est le formalisme g\'en\'eral dont ils sont tir\'es qui lui en assure la
coh\'erence interne, sans qu'il soit besoin de faire appel \`a une
hypoth\'etique th\'eorie hilbertienne.
\section{Evolutions}
Pour fixer les id\'ees, consid\`erons une particule quantique sans spin
 d\'ecrite par la valeur de $t$ d\'efinissant l'instant actuel et par
la fonction
$\phi_t(\vec x)$ d\'efinissant \`a cet instant ses propri\'et\'es actuelles.
Deux cas extr\^emes sont relativement faciles \`a d\'ecrire.
a){\bf L'\'evolution d\'eterministe et r\'eversible}: Pour ce cas on
d\'emontre tout d'abord
que la relation
d'or\-thogonalit\'e doit \^etre conserv\'ee et donc, que l'\'evolution
est induite par une tranformation unitaire. Ainsi $\phi_t(\vec x)$
ob\'eit \`a une \'equation de Schr\"odinger:
$$i\partial_t\phi_t(\vec x)=H\phi_t(\vec x)$$
o\`u $H$ est un op\'erateur auto-adjoint qui peut d\'ependre du temps et
dont la forme est impos\'ee par le principe de Galil\'ee.
C'est l\`a un r\'esultat important qui se d\'emontre
[Piron 1998 3.3.1, 3.3.5] et n'a donc pas besoin d'\^etre postul\'e.
b){\bf Le processus de mesure id\'eale}: Parmis les projets exp\'erimentaux
qui d\'efinissent une propri\'et\'e donn\'ee $a$, il y en a qui
perturbent compl\`etement le syst\`eme ou m\^eme
 le d\'etruisent. Mais il y en a aussi qui le perturbent le moins
possible, fournissent une information et sont en quelques sortes
des mesures. C'est pourquoi on d\'efinira une mesure id\'eale $\alpha$
par les trois conditions suivantes [Piron 1998 3.5]:
1)$\alpha$ est un projet exp\'erimental d\'efinissant $a$, dont la r\'eponse
positive est impossible si $a^\perp$ est actuelle.
2)En cas de r\'eponse positive, la propri\'et\'e $a$ est actuelle en
fin de mesure.
3)De plus dans ce cas, le syst\`eme a \'et\'e perturb\'e le moins possible.
On montre facilement que si la r\'eponse est positive  l'effet d'une telle
mesure est de transformer l'\'etat initial $\phi_{t_1}$ dans
l'\'etat final $\phi_{t_2}=P_a\phi_{t_1}$ o\`u $P_a$ est le projecteur
associ\'e \`a $a$. La fonction $P_a\phi_{t_1}$ n'est pas
forc\'ement normalis\'ee mais elle n'est jamais nulle en vertu de la premi\`ere
condition.
Il est possible de d\'eterminer pour ce type de mesure la probabilit\'e
a priori d'obtenir une r\'eponse positive, c'est-\`a-dire ind\'ependente des
conditions particuli\`eres du d\'eroulement de l'exp\'erience. Notons
$w(a,\phi)$ cette probabilit\'e et supposons qu'elle existe
pour tout $a$ et tout $\phi$. Vu son interpr\'etation, il est naturel de
lui imposer les trois conditions suivantes:
1)$w(a,\phi)=1$, si $a$ est actuelle pour $\phi$.
2)$w(a^\perp,\phi) = 1- w(a,\phi)$, car $a$ et $a^\perp$ correspondent
souvent \`a la m\^eme exp\'erience.
3)$w(a\cap b,\phi) = w(a,\phi)w(b,P_a\phi)$ chaque fois que $P_a$ et
$P_b$ commutent. Car dans ce cas la mesure id\'eale de $a$ suivie de
la mesure id\'eale de $b$ peut \^etre interpr\'et\'ee comme une mesure
id\'eale de $a\cap b$. Pour s'en convaincre il faut tout d'abord remarquer
que dans le cas d'une double r\'eponse
positive l'\'etat final sera bien $P_{a\cap b}\phi = P_bP_a\phi$.
En suite il faut montrer que si l'\'etat initial est dans
$(a\cap b)^\perp$ c'est-\`a-dire, de la forme $(I-P_bP_a)\phi$, alors
une double r\'eponse positive est impossible. Or si  la premi\`ere
r\'eponse a \'et\'e positive,
l'\'etat avant la seconde mesure sera de la forme $(I-P_b)P_a\phi$ et
une deuxi\`eme r\'eponse positive est alors impossible.
Il est tout \`a fait remarquable qu'une telle fonction existe et
 que de plus elle soit unique:
$$w(a,\phi) = \Vert\phi\Vert^{-2}\Vert P_a\phi\Vert^2$$
L'unicit\'e de cette fonction en montre l'universalit\'e et explique
 pourquoi elle est
tr\`es souvent appliqu\'ee. Il est important de remarquer que nous n'avons
utilis\'e pour sa d\'erivation que les lois habituelles des probabilit\'es,
ce qui montre \`a l'\'evidence que cette probabilit\'e est
de m\^eme nature que celles qui interviennent en th\'eorie classiques.
Enfin cette formule est l'\'equivalent en th\'eorie quantique des
probabilit\'es \'egales a priori. Il n'y a pas ici de probl\`eme
d'interpr\'etation de la mesure et de r\'eduction du paquet d'ondes
car la d\'efinition de l'\'etat n'utilise pas ces concepts et c'est bel
et bien l'appareil qui perturbe le syst\`eme et non les pens\'ees de
l'observateur.
Mesurer une observable $A$ d\'efinie par sa d\'ecomposition spectrale
$$A=\sum_{\scriptstyle i\in J}\lambda_iP_{a_i}$$
revient \`a effectuer une s\'erie de mesures id\'eales $P_{a_i}$. Vu
 la condition 3) on peut calculer directement et justifier ainsi la
formule bien connue:
$$\bar{\lambda} = \sum_{\scriptstyle i\in J}\lambda_i
\Vert\phi\Vert^{-2}\Vert P_{a_i}\phi\Vert^2
    = tr(PA)$$
o\`u $P$ d\'esigne le projecteur sur l'\'etat $\phi$ et $\bar\lambda$
la moyenne statistique des $\lambda_i$ ainsi obtenus.
Souvent l'\'etat du syst\`eme mesur\'e n'est pas compl\`etement connu,
tout au plus peut-on supposer donn\'e un ensemble $E$ d'\'etats
$\phi_\omega$ et une distribution de probabilit\'e $d\mu(\omega)$
[Cohen-Tannoudji 1973 E$_{III}$.4.a. et Piron 1998 3.5.6]
Dans ce cas la formule pr\'ec\'edente se g\'en\'eralise en
rempla\c cant le projecteur $P$ par l'op\'erateur
$$\rho = \int_EP_\omega d\mu(\omega)$$
o\`u $P_\omega$ d\'esigne le projecteur sur $\phi_\omega$. L'op\'erateur
$\rho$ est appel\'e la matrice densit\'e. Par construction la donn\'ee
de $\rho$ est suffisante pour
calculer toutes les valeurs moyennes de ce type. Pour d'autres
exp\'eriences, o\`u interviennent explicitement les $\phi_\omega$ et leurs
corr\'elations, la donn\'ee de $\rho$ ne suffit cependant pas
pour pr\'edire
les r\'esultats [Cohen-Tannoudji 1973 E$_{III}$.4.c. et Piron 1998 3.6.5].
c)En plus de ces deux types d'\'evolutions, il
y a des cas interm\'ediaires. Nous nous contenterons d'un
exemple: Un spin $\frac12$ plac\'e dans un champ magn\'etique homog\`ene
tourne autour de l'axe de ce champ, c'est la pr\'ecession de Larmor
que d\'ecrit une transformation unitaire. Mais en m\^eme temps, en
r\'ealit\'e, l'angle $\theta$ que fait le spin avec la direction du
champ diminue lentement, c'est la relaxation longitudinale, le syst\'eme
tend ainsi vers un \'equilibre o\`u $\theta$ est nul.
Le tout est encore un processus d\'eterministe mais qui ne peut alors
\^etre d\'ecrit que
par une \'equation non-lin\'eaire, par exemple:
$$i\partial_t\phi_t=-\gamma\frac12\vec B\vec\sigma\phi_t
 +i(T_1\vert\vec B\vert)^{-1}\frac12(\vec B\vec\sigma
-<\phi_t,\vec B\vec\sigma\phi_t>)\phi_t $$
Le premier terme d\'ecrit la pr\'ecession de Larmor de pulsation
$\omega_L= -\gamma\vert\vec B\vert$ et le second une
 relaxation longitudinale de temps caract\'eristique
$T_1$.
 Il ne faudrait pas prendre trop au s\'erieux cette  \'equation
qui ne d\'ecrit que l'\'evolution d\'eterministe
d'un spin seul, en ignorant l'action al\'eatoire du reste du syst\`eme
 qui lui aussi contribue \`a la relaxation.
Afin de d\'emontrer sur cet exemple particulier, que
conform\'ement \`a la th\'eorie g\'enerale que nous avons donn\'ee au
tout d\'ebut, il est parfaitement possible de
d\'ecrire une th\'eorie quantique uniquement en termes
d'\'el\'ements de r\'ealit\'e, transcrivons dans ces
termes l'\'equa\-tion d'\'evolution pr\'ec\'edente. Un \'etat quelconque
de spin $\frac12$ peut s'\'ecrire:
$$\cos(\frac12\theta)\exp(-i\frac12\varphi)
\vert+> +\, \sin(\frac12\theta)\exp(i\frac12\varphi)\vert->$$
o\`u $\vert+>$ et $\vert->$ d\'esignent les vecteurs propres du spin
dans la direction choisie, celle du champ magn\'etique [Cohen-Tannoudji
1973 IV A.2.b. et Piron 1998 2.9.2]. Or cet \'etat, not\'e
$\vert\theta,\varphi>$, d\'esigne un
\'el\'ement de r\'ealit\'e: un spin
$\frac12$ pointant dans la direction $(\theta,\varphi)$ par rapport
au champ magn\'etique.
Il n'est alors pas trop difficile de v\'erifier
que l'\'equation non-lin\'eaire pr\'ec\'edente est bien
\'equivalente aux deux \'equations suivantes:
$$\dot\theta={-\sin\theta\over T_1}\phantom{aussi}
 {\rm et}\phantom{aussi} \dot\varphi=\omega_L$$
C'est l\`a un cas tr\`es simple, d\^u au fait que l'espace de Hilbert
du spin $\frac12$ est de dimension deux. Il suffit de
 deux nombres pour d\'efinir l'\'etat \`a un instant donn\'e, alors
qu'en m\'ecanique il en faut  six pour rep\`erer l'\'etat d'une
particule classique et  une infinit\'e pour une particule quantique.
Pour terminer et illustrer ces concepts nous allons discuter
l'exp\'erience des \'echos de spins. Si sur un \'echantillon d'eau
on applique un fort
champ magn\'etique dans la direction verticale et si ensuite on
supprime brutalement ce champ, les spins des protons qui s'\'etaient
align\'es dans la direction verticale vont se mettre tous ensemble
\`a pr\'ecesser librement autour de l'axe $z$ du champ r\'esiduel,
le champ magn\'etique terrestre. Le syst\`eme ainsi obtenu peut
valablement \^etre
d\'ecrit par la matrice densit\'e \`a un corps:
$$\rho=\vert\theta,\varphi><\theta,\varphi\vert={1\over2}\left(\matrix{
1+\cos\theta&\exp(-i\varphi)\sin\theta\cr
-\exp(+i\varphi)\sin\theta&1-\cos\theta\cr}\right)$$
o\`u comme pr\'ec\'edemment
$$\vert\theta,\varphi>=\cos(\frac12\theta)\exp(-i\frac12\varphi)
\vert+> +\, \sin(\frac12\theta)\exp(i\frac12\varphi)\vert->$$
Ce mouvement se d\'etecte facilement avec une bobine d'axe horizontal
perpendiculaire
\`a $z$, mais tr\`es vite le signal dispara\^{\i}t car chaque spin
voit un champ l\'eg\`erement diff\'erent et l'angle de rotation
d\^u \`a la pr\'ecession de Larmor devient al\'eatoirement diff\'erent
d'un spin \`a l'autre. La nouvelle matrice densit\'e de l'ensemble
est alors donn\'ee par:
$$\rho=\int_E\vert\theta,\varphi><\theta,\varphi\vert d\varphi={1\over2}
\left(\matrix{1+\cos\theta&0\cr
0&1-\cos\theta\cr}\right)$$
Ainsi les termes hors-diagonales, les termes de coh\'erence, ont disparu.
On serait tent\'e de croire
qu'il n'est donc pas possible de distinguer cet ensemble d'un
m\'elange pond\'er\'e de spins $\vert+>$ et $\vert->$
[Cohen-Tannoudji E$_{IV}$.3. 1973].
Cette conclusion serait parfaitement valable si on se limitait
exp\'erimentalement \`a des mesures id\'eales de spins faites
individuellement et au hasard. Mais
l'exp\'erimentateur peut agir autrement. En envoyant au temps $t$,
perpendiculairement \`a $z$ et dans le plan verticale de d\'epart,
une forte et courte impulsion de champs magn\'etique correspondant \`a un
angle de Larmor de $\pi$ les spins apr\`es avoir bascul\'es
continuent de tourner chacuns de la m\^eme fa\c con, mais comme
les lents se sont retouv\'es en t\^ete et les rapides en queue
 ils se retrouvent tous en phase au temps $2t$  et le signale
r\'eappara\^{\i}t. C'est le ph\'enom\`ene d'\'echos de spins [Hahn 1950],
en r\'ep\'etant cette impulsion au temps $3t$ on obtient un second
\'echo au temps $4t$ et ainsi de suite.
 Ces \'echos d\'ecroissent avec le temps, mais il faut remarquer que
ce type de relaxation n'est pas contenue dans l'\'equation du mouvement
pr\'ec\'edente car l'effet du terme en $T_1$, \`a cause du retournement,
s'exerce une fois dans un sens et une fois dans l'autre et cela pendant
le m\^eme temps. Il n'en serait pas de m\^eme, si on obtenait ces
\'echos sans
retouner les spins, en inversant le gradient d'un petit champs
magn\'etique ajout\'e au champs terrestre tr\`es homog\`ene [Borcard 1972].
En retournant alors ces spins par une impusion de $\pi$ on peut m\^eme esp\'erer
r\'eg\'en\'erer l'\'echo.

\references
\parindent=0pt
\def\refjl#1#2#3#4#5{\hangafter=1\hangindent=30pt
#1; ``#2'' {\it #3} {\bf #4} #5.\par}
\def\refbk#1#2#3{\hangafter=1\hangindent=30pt
#1; ``#2'' #3.\par}
\def\refpr#1#2#3#4#5{\hangafter=1\hangindent=30pt
#1; ``#2'' in #3 {\it #4} #5.\par}
\refjl{A. Einstein, B. Podolsky, N. Rosen 1935}{Can quantum-mechanical
description of reality be considered complete?}{Phys. Rew.}{47}{777-780}
\refjl{E.L. Hahn 1950}{Spin Echoes}{Phys.Rew.}{80}{580-594}
\refjl{C. Piron 1965}{Sur la quantification du syst\`eme de deux
particules}{Hel. Phys. Acta}{38}{104-108}
\refbk{C. Cohen-Tannoudji, B. Diu, F. Lalo\"e 1973}{M\'ecanique quantique}
{Hermann, Paris}
\refjl{B. Borcard et G.J. B\'en\'e 1972}{Echos de spins par inversion d'un
gradient de champ magn\'etique}{Hel.Phys.Acta}{45}{64-66}
\refjl{D. Aerts 1982}{Description of many separated physical entities without
the paradoxes encountered in quantum mechanics}{Found. Phys.}{12}{1131-1170}
\refpr{C. Piron 1990}{Time, Relativity and Quantum Theory}{New Frontiers
in Quantum Electrodynamics and Quantum Optics}{Edited by A. O. Barut ,
Plenum Press, New York}{495-506}
\refjl{A. O. Barut, D. J. Moore and C. Piron 1993}{The Cartan formalism in
field theory}{Helv. Phys. Acta}{66}{795-809}
\refjl{C. Piron and D. J. Moore 1995}{New aspects of field theory}{Turk. J.
Phys.}{19}{202-216}
\refjl{C. Piron 1996}{Quantum Theory without Quantification}{Helv. Phys.
Acta}{69}{694-701}
\refbk{C. Piron 1998}{M\'ecanique quantique bases et applications}
{Presses polytechniques et universitaires romandes, Lausanne}
\refbk{Claude Audoin et Bernard Guinot 1998}{Les fondements de la
mesure du temps, comment les fr\'equences atomiques r\`eglent le monde}
{Masson, Paris}
\bye